\documentclass[10pt,a4paper]{article}
\usepackage{amsmath}
\usepackage{amsthm}
\usepackage{amsfonts}
\usepackage{multicol}
\usepackage[dvips]{graphicx}
\usepackage{fixltx2e}

\newcommand{\danger}[1]{\textbf{#1}}

\input{epsf}             % encapsulated Postscript figures
\begin{document}

\title{\danger{2D Naked Singularity in General Relativity}}
\author{\centerline{\danger{J. Manuel Garc\'\i a-Islas \footnote{
e-mail: jmgislas@iimas.unam.mx}}}  \\
Departamento de F\'\i sica Matem\'atica \\
Instituto de Investigaciones en Matem\'aticas Aplicadas y en Sistemas \\ 
Universidad Nacional Aut\'onoma de M\'exico, UNAM \\
A. Postal 20-726, 01000, M\'exico DF, M\'exico\\}

\maketitle

\begin{abstract}
We present a novel example of a 2-dimensional space-time naked singularity. The solution
has a gravity singularity and no-horizon. 
This example is only a toy model and as such its motivation is mathematical. In the physical sense  
it is very helpful to understand in simple terms what naked singularities are
and the properties they may have.  
\end{abstract}

\section{Introduction}

A naked singularity or gravitational singularity can be observed from infinity. 
It has been stated as the Cosmic Censorship Conjecture \cite{rp} 
that the collapse 
of a generic and physically realistic star in general relativity
would never lead to a space-time singularity which is visible to faraway observers. 
All singularities 
formed from the collapse are always behind a curtain known as the event horizon 
and hence invisible to outside observers. Many mathematical 
attempts have been made to prove the Cosmic Censorship Conjecture. 
On the other hand, there is a rigorous mathematical proof of the formation of black hole 
singularities in general relativity \cite{hp}.

Naked singularities, for example, appear in four dimensions in the Reissner-Nordstr\"om metric \cite{hel}.
They have certain properties and have been studied extensively. For example, an ingoing particle in the 
Reissner-Nordstr\"om metric is repelled 
in the vicinity of the naked singularity at a minimum $r=r_{min}$, 
an accelerates back as an outgoing particle \cite{pjkn}.  
Besides Reissner-Nordstr\"om, naked singularities have been studied in the  
literature quite a lot, from the classical point of view to the quantum one \cite{e}.

The example we present here has the peculiarity that it is surprisingly simple, completely new to our
knowledge, and with similar properties 
to Reissner-Nordstr\"om but also with important differences. For instance, in our toy model there is only gravity (or curvature)
and no charge, that is, nothing which gives place to another field. We will also have the property that
ingoing particles are repelled 
in the vicinity of the naked singularity at a minimum $r=r_{min}$, 
an accelerate back as outgoing ones.  

The following paper is divided as follows: 
In section 2 we introduce the space-time naked singularity. In section 3 we study its properties.
In section 4 we change coordinates to Kruskal like ones and construct its Penrose Diagram in order
to understand a bit deeper this naked singularity.  
In section 5 we conclude with a brief discussion on the similarities and differences to Reissner-Nordstr\"om 
geometry and to $1+3$ general relativity in general.

\section{Naked Singularity}

The example we consider is not a real physical naked singularity since it is 
a $2$-dimensional case\footnote{In relativity notation it is a 1+1 solution.}. It is only a toy model, and even if it is a simple example 
or a starting point if the reader will, 
it is a very helpful solution which helps a lot in order
to understand naked singularities in $1+3$-dimensions.

Consider Einsten's equations of general relativity

\begin{equation}
R_{\mu \nu} - \frac{1}{2} R g_{\mu \nu} = 8 \pi G T_{\mu \nu}
\end{equation}
It is known that in
two dimensions equation $(1)$ is satisfied identically\footnote{This means that any arbitrary 2-dimensional metric
satisfies the 2-dimensional Einstein's equations. However, this does not mean that any metric
is interesting for relativity. 

The one we present here is really an interesting one for relativity since
it has a naked singularity.} \cite{adb}.
That means that the
energy-momentum tensor always vanishes. 
Because of this the metric is not constrained.

Consider the following metric\footnote{Throughout this paper we are using units where the speed of light is equal to one.}

\begin{equation}
ds^2 = - \coth r \ dt^2 + \tanh r \ dr^2
\end{equation}
The metric has the property that it is defined for $- \infty < t < \infty$ and for
$0 < r < \infty$ as well as for $- \infty < r < 0$.\footnote{Observe that for this latter case the metric changes signature.
The time coordinate turns into a space coordinate and vice versa.} 
As we shall see there is a curvature singularity at $r=0$.\footnote{The metric is undefined at $r=0$.} 

Let $0 < r < \infty$.

The metric approaches the $2$-dimensional Minkowskian one when $r \rightarrow \infty$.

It has Christoffel symbols, curvature coefficients,
Ricci tensors and scalar curvature given by

\begin{align}
\Gamma_{rt}^{t} = - \ \frac{1}{\sinh 2r} \ \ \ , \ \ \  \Gamma_{rr}^{r} = \frac{1}{\sinh 2r}
\ \ \ , \ \ \ \Gamma_{tt}^{r} = - \frac{1}{2} \ \frac{\coth r}{\sinh^2 r} \nonumber
\end{align}

\begin{align}
R_{rtr}^{t} = - \ \frac{1}{\sinh^2 r} \ \ \ , \ \ \  R_{trt}^{r} =  \ \frac{\coth^2 r}{\sinh^2 r} \nonumber
\end{align}

\begin{align}
R_{tt} = \frac{\coth^2 r}{\sinh^2 r}  \ \ \ , \ \ \  R_{rr} = - \ \frac{1}{\sinh^2 r} \nonumber
\end{align}

\begin{equation}
R = - \ 2 \ \frac{\coth r}{\sinh^2 r} \nonumber
\end{equation}
The scalar curvature is completely negative and it satisfies, 
$R \rightarrow - \infty$ when $r \rightarrow 0$ and $R \rightarrow 0$ when 
$r \rightarrow \infty$. See Figure 1. 

Observe that the scalar curvature is almost flat for $r>2$, but changes drastically when 
the radial coordinate is smaller than 2. 
The scalar curvature is undefined only at $r=0$ and there is no horizon in this geometry.

\begin{figure}[h]
\begin{center}
\includegraphics[width=0.7\textwidth]{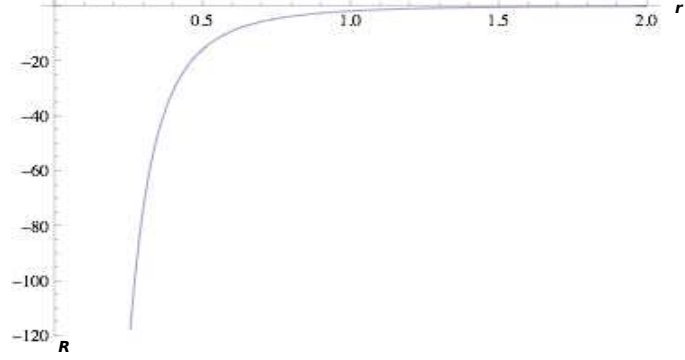}
\caption{Scalar curvature function $R$.}
\end{center}
\end{figure}

\begin{figure}[h]
\begin{center}
\includegraphics[width=0.8\textwidth]{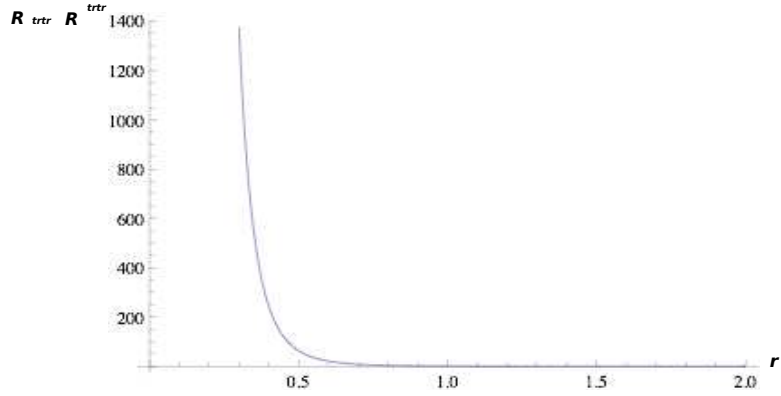}
\caption{Kretschmann scalar $R_{t r t r} R^{t r t r}$}
\end{center}
\end{figure}

The Kretschmann scalar $R_{\mu \nu \sigma \rho}R^{\mu \nu \sigma \rho}$ can be calculated easily.
It is known that the curvature tensor in $n$-dimensions has $\frac{n^2 (n^2-1)}{12}$ independent components 
Therefore it has only one component in two dimensions which in the present case it is given by

\begin{equation}
R_{t r t r} =  \frac{\coth r}{\sinh^2 r} 
\end{equation}
and the Kretschmann scalar by

\begin{equation}
R_{t r t r} R^{t r t r}=  \frac{\coth^2 r}{\sinh^4 r} 
\end{equation}

The Kretschmann scalar is positive and it satisfies, 
$R_{t r t r} R^{t r t r} \rightarrow \infty$ when $r \rightarrow 0$ and $R_{t r t r} R^{t r t r} \rightarrow 0$ when 
$r \rightarrow \infty$. See Figure 2.

Because of the behaviour of the scalar curvature and the Kretschmann scalar we have a naked singularity.

\section{Particles at the naked singularity geometry}

Consider first the motion of photons in this geometry:

\begin{equation}
- \coth r \ dt^2 + \tanh r \ dr^2 = 0
\end{equation}
from which it is obtained

\begin{equation}
\frac{dr}{dt} = \pm \coth r
\end{equation}
We have the outgoing and ingoing null curves given by

\begin{equation}
t = \pm \int \tanh r \ dr = \pm \log[\cosh r] + a
\end{equation}
from which we can compute the time\footnote{Coordinate time which is the time a faraway observer measures} 
a photon takes to go from a coordinate
radius to another one. $a$ is a constant.
It can be seen that ingoing null curves have no problem to reach 
the singularity $r=0$ in finite time $t$. See Figure 3.

\begin{figure}[h]
\begin{center}
\includegraphics[width=0.9\textwidth]{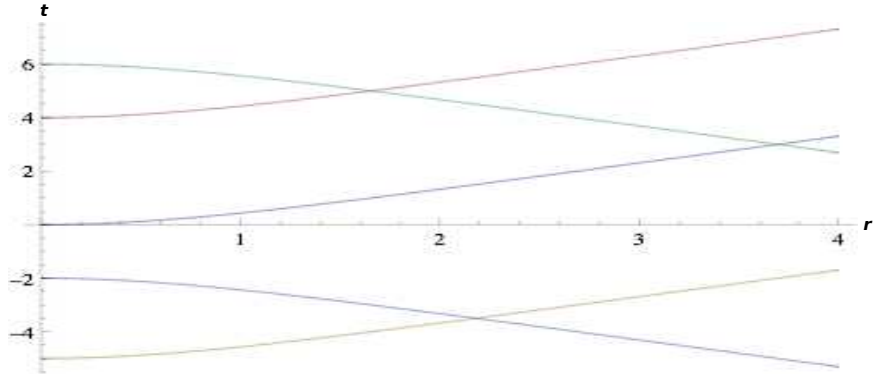}
\caption{Null curves at the naked singularity geometry}
\end{center}
\end{figure}

Consider massive particles.
The proper time of a stationary particle at a fixed $r$ and of a far away particle are related by

\begin{equation}
\Delta \tau = ( \coth r )^{1/2} \ \Delta t
\end{equation}
It can be seen that the proper times coincide when $r \rightarrow \infty$. The proper time
of a massive particle approaching the singularity runs faster, and tends to infinity, in opposition to Schwarzschild
space-time for example. However, as a massive particle will not be able to reach the singularity, it will take a finite
proper time for the particle to reach the minimum radial coordinate allowed, as we will see.
Because of this behaviour there is a gravitational blueshift, since if a photon is sent from $r_E$ to
a receptor $r_R$, where $r_E < r_R$, it can be seen that $\nu_{R} > \nu_{E}$.

If we consider the motion of massive particles, we have

\begin{equation}
- \coth r \ \bigg(\frac{dt}{d\tau}\bigg)^2 + \tanh r \ \bigg(\frac{dr}{d\tau}\bigg)^2 = -1
\end{equation}
and the geodesic equation corresponding to the $t$ coordinate is given by

\begin{equation}
\coth r \ \bigg(\frac{dt}{d\tau}\bigg) = K
\end{equation}
$K$ is naturally interpreted as the energy measured by the observer at rest located at infinity.   
We substitute equation $(10)$ into equation $(9)$ and get

\begin{equation}
\bigg(\frac{dr}{d\tau}\bigg)^2 + \coth r= K^2
\end{equation}
From this equation we can observe that for a massive particle with initial conditions
$\frac{dr}{d\tau} = 0$ 
at $r=R$, we will have $K^2 = \coth R$ and it will imply

\begin{equation}
\bigg(\frac{dr}{d\tau}\bigg)^2 = \coth R - \coth r
\end{equation}
where the right side is negative for decreasing $r$ $(r<R)$ and positive for increasing $r$ $(r > R)$. 
This means that a
massive particle at rest at a fixed radial coordinate $R$, will move upwards, away from the singularity.
However, it can be seen from formula $(12)$, that it will remain almost at rest and will barely move,
unless $R$ is really small. 

Equation $(12)$ implies that if the massive particle
were able to put itself very close to the singularity $R \sim 0$ the particle will move upwards
and will be able to reach larger values of $r$ at a speed measured by a fixed observer at $r$ given by    

\begin{equation}
v' = \frac{dR'}{dt'} = \frac{\sqrt{\tanh r}}{\sqrt{\coth r}}\frac{dr}{dt} = \frac{1}{\coth r} \frac{dr}{dt} = \sqrt{1- \frac{\coth r}{K^2}}
\end{equation}
where $t'$ is the proper time of the observer at fixed radial coordinate $R'$.
It is easily seen that for an observer fixed at infinity and when energy goes to infinity $v' \rightarrow 1$, which is the
speed of light.

As discussed before, massive particles at rest away from the singularity will barely move, 
and in order to approach the singularity they will have to move with initial velocity different to zero.

If a massive particle has initial velocity $\frac{dr}{d\tau} = v$, where $v <1$ at an initial position $r=R$, it
will be able to move close to the singularity. Because of the initial condition, equation $(11)$ implies 
that $K^2 = v^2 + \coth R$. This means that

\begin{equation}
\bigg(\frac{dr}{d\tau}\bigg)^2 = v^2 + \coth R - \coth r
\end{equation}

\begin{figure}[h]
\begin{center}
\includegraphics[width=0.8\textwidth]{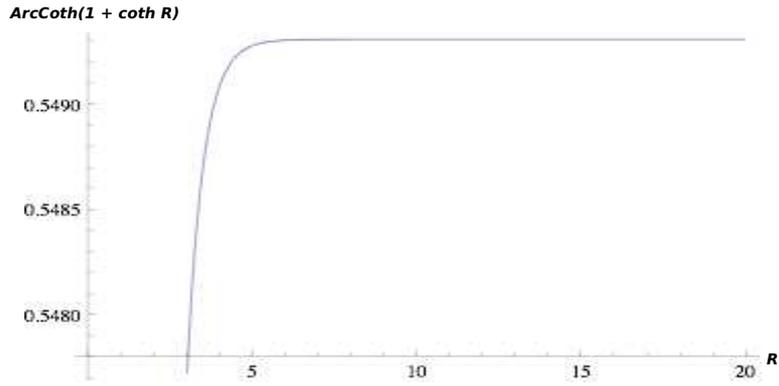}
\caption{$r= \coth^{-1}(1+\coth R)$}
\end{center}
\end{figure}

As this squared velocity can be positive it means that the massive particle can move
downwards to the singularity. 
It can be seen that if a massive particle at radial coordinate 
$r=R$ has an initial velocity 
close to that of light $v \sim 1$, we will have from equation $(12)$ that the particle will 
inevitably go to rest at $r \sim \coth^{-1}(1 + \coth R)$.

This can be seen 
from the graph of the function $r= \coth^{-1}(1 + \coth R)$ plotted in Figure 4.
A massive particle which starts moving from $R$ with initial velocity very close to that
of light, will at most reach the value $r \sim 0.5493$ regardless of the value of $R$.

Once the particle is at this latter position, it will be at rest and then 
it will go upwards really fast, and will approach the velocity of
light without surpassing it.    

A massive particle will be able to approach the naked singularity but will not be able to fade away
from this universe by hitting the singularity.

 \section{Kruskal coordinates and Penrose Diagram}
 
 In this section we change the naked singularity metric to Kruskal like coordinates and construct
 its Penrose diagram. The purpose of this is to understand deeper how this naked singularity behaves.
 
 The construction is just straightforward since the procedure is analogous to the Schwarzschild one for example.
 
 The ingoing null curves are given by equation $(7)$ 
 
\begin{equation}
t = -  \log[\cosh r] + p
\end{equation}
where $p$ is a constant. Using this constant as a new coordinate we have     

\begin{equation}
p = t + \log[\cosh r] 
\end{equation}
and in terms of this coordinate the metric $(2)$ takes the form

\begin{equation}
ds^2 = - \coth r \ dp^2 + 2 dp  \ dr
\end{equation}
In terms of these coordinates the ingoing and outgoing null curves $ds^2=0$, are given respectively by

\begin{figure}[h]
\begin{center}
\includegraphics[width=0.8\textwidth]{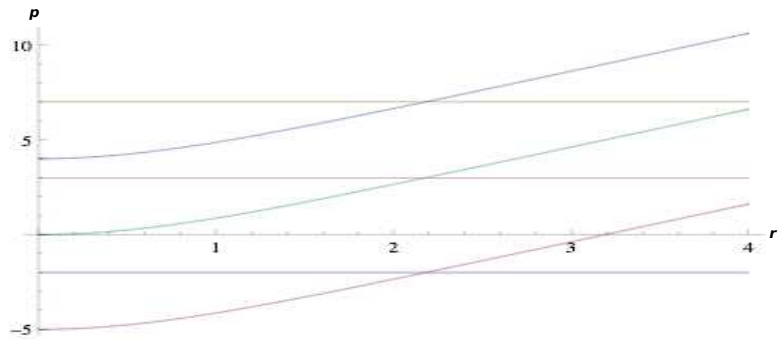}
\caption{Null curves in the $(p,r)$ coordinates.}
\end{center}
\end{figure}

\begin{equation}
p = \alpha \nonumber 
\end{equation}
\begin{equation}
p= 2 \log[\cosh(r)] + \beta
\end{equation}
where $\alpha$ and $\beta$ are constants. See Figure 5.

Likewise if we had chosen the outgoing null curves from equation $(7)$

\begin{equation}
t = + \log[\cosh r] + q
\end{equation}
where $q$ is a constant. Using this constant as a new coordinate we have     

\begin{equation}
q = t - \log[\cosh r] 
\end{equation}
and in terms of this coordinate the metric $(2)$ takes the form

\begin{equation}
ds^2 = - \coth r \ dq^2 - 2 dq \ dr
\end{equation}
In terms of these coordinates the outgoing and intgoing null curves $ds^2=0$, are given respectively by

\begin{figure}[h]
\begin{center}
\includegraphics[width=0.8\textwidth]{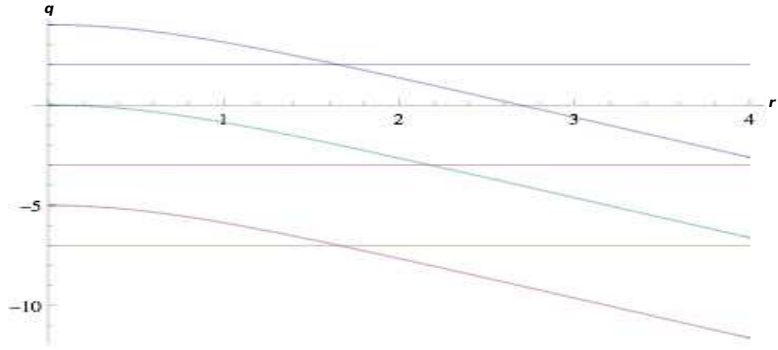}
\caption{Null curves in the $(q,r)$ coordinates.}
\end{center}
\end{figure}

\begin{equation}
q = \alpha \nonumber 
\end{equation}
\begin{equation}
q = - 2 \log[\cosh(r)] + \beta
\end{equation}
where $\alpha$ and $\beta$ are constants. See Figure 6.
In terms of $(p,q)$ coordinates the metric $(2)$ becomes

\begin{equation}
ds^2 = - \coth r \ dp  \ dq
\end{equation}
where $r$ is given in terms of $p$ and $q$ by

\begin{equation}
\frac{1}{2}(p-q) = \log[\cosh r] 
\end{equation}
Let us now consider the Kruskal like change of variables given for convenience by

\begin{equation}
\tilde{p}= \exp\bigg(\frac{p}{2}\bigg) \ \ \ \ \ \ \ \ \tilde{q}= - \exp\bigg(\frac{-q}{2}\bigg)
\end{equation}
In $(\tilde{p}, \tilde{q})$ coordinates the naked singularity metric is given by

\begin{equation}
ds^2 = - \frac{4}{\sinh r} \ d\tilde{p}  \ d\tilde{q}
\end{equation}
If we now take a time-like variable $\tilde{t}$ and a space-like variable $\tilde{r}$ defined by

\begin{equation}
\tilde{t}= \frac{1}{2} ( \tilde{p} + \tilde{q} ) \ \ \ \ \ \ \ \ \tilde{r}= \frac{1}{2} ( \tilde{p} - \tilde{q} )
\end{equation}
the naked singularity metric is given in Kruskal like coordinates by

\begin{equation}
ds^2 = \frac{4}{\sinh r} \ ( - \ d\tilde{t}^2 + d\tilde{r}^2 )
\end{equation}
where it is easily seen that $\tilde{t}$ and $\tilde{r}$ are expressed in terms of the original $t$ and $r$ by the transformations
  
\begin{equation}
\tilde{t}= (\cosh r)^{1/2} \sinh \frac{t}{2} \ \ \ \ \ \ \ \ \tilde{r}= (\cosh r)^{1/2} \cosh \frac{t}{2}
\end{equation}  
and therefore

\begin{equation}
- \tilde{t}^2 + \tilde{r}^2 = \cosh r \ \ \ \ \ \ \ \ \frac{\tilde{t}}{\tilde{r}}= \tanh \frac{t}{2}
\end{equation} 

\begin{figure}[h]
\begin{center}
\includegraphics[width=0.8\textwidth]{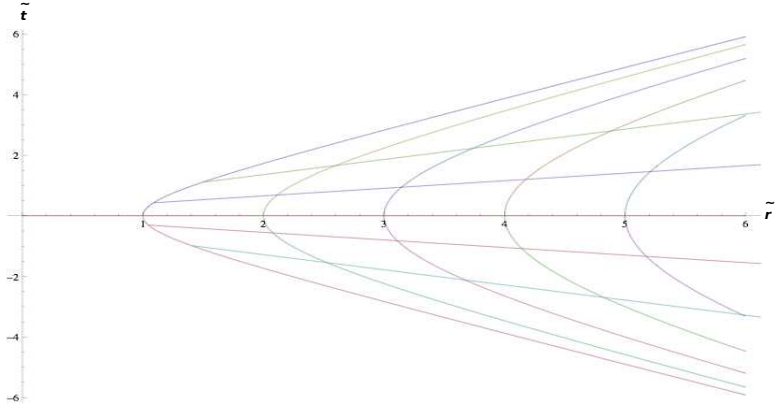}
\caption{Kruskal like coordinates $(\tilde{t} , \tilde{r})$ of the naked singularity.}
\end{center}
\end{figure}
In Figure 7 we can observe the Kruskal like coordinates of the naked singularity. They have analogous properties 
to the Schwarzschild case, but they also have important differences. From the domain of the original set of coordinates   
$- \infty < t < \infty$ and $0 < r < \infty$ we notice that the Kruskal like coordinates $(\tilde{t} , \tilde{r})$ must take values
$- \infty < \tilde{t} < \infty$ and $1 < \tilde{r} < \infty$. Moreover we have that $- \tilde{t}^2 + \tilde{r}^2 > 1$.

This means for instance that the line singularity $r=0$ is given by the hyperbola $- \tilde{t}^2 + \tilde{r}^2=1$ in the 
Kruskal like coordinates. The set of hyperbolas $- \tilde{t}^2 + \tilde{r}^2 = \cosh r$ for $0 < r < \infty$ represent
lines of constant $r$, that is, fixed observers. Likewise lines of constant $t$ are given by the lines  
$\frac{\tilde{t}}{\tilde{r}}= \tanh \frac{t}{2}$. In Figure 7 lines of constant $r$ and of constant $t$ are drawn. 

Past infinity and future infinity are given by the asymptotes of the whole set of hyperbolas. 

\begin{figure}[h]
\begin{center}
\includegraphics[width=0.5\textwidth]{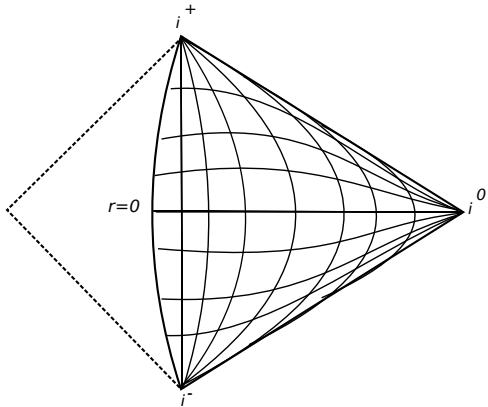}
\caption{Penrose diagram of the naked singularity.}
\end{center}
\end{figure}

By considering the transformations 

\begin{equation}
p ' = \tanh \tilde{p} \ \ \ \ \ \ \ \  q ' = \tanh \tilde{q}
\end{equation}
the Penrose diagram can be drawn as in Figure 8.

By examining the Kruskal like coordinates and the Penrose diagram some of the properties of
the naked singularity discussed in the previous section can be observed.

We mention at the beginning of section 2, that the naked singularity solution is also valid for
negative values of $r$, $- \infty < r < 0$. In this latter case space and time are interchanged and
the signature of the metric changes as well. This is equivalent to considering the negative of metric $(2)$
and $r$ positive definite.
Therefore it can be seen that the corresponding Kruskal
and Penrose diagram for this case is equivalent to rotating Figure 7 and Figure 8 by
90 degrees leaving axis $\tilde{t}$ and $\tilde{r}$ fixed in Figure 7.

\section{Conclusion}

We have introduced a very simple example of a $1+1$ naked singularity geometry and described its
properties. There is an important difference to general solutions in $1+3$ general relativity. For instance,
we have seen, in the present example that gravity is not attractive, particles at rest will not fall. 
On the contrary, they almost remain motionless and will move upwards really slowly. 
This is due to the fact that the solution is almost
Minkowskian at almost any radial coordinate $r$, and only when the radial coordinate is small 
there are important issues. These issues are that massive particles are really affected by the geometry of
space-time but in an antigravity way, as particles are repelled by the singularity and start to move
really fast to larger values of $r$. This resembles what happens in the Reissner-Nordstr\"om
solution and particularly in the naked singularity case when $M^2 < Q^2$ studied in \cite{pjkn}.
The difference is that in the Reissner-Nordstr\"om case particles fall to the singularity whereas in the
present example particles will have to move on their own towards the singularity and then they will be repelled.

We also have that 
in $1+3$ general relativity, for example in Schwarzschild geometry, there is a gravitational
redshift when photons are emitted at $r_E$ and received at $r_R$, where $r_E < r_R$; 
whereas here we have a blueshift.

In addition, we considered the Kruskal like coordinates and the Penrose diagram of the naked
singularity. The change to that kind of coordinates is straightforward and it is interesting to notice
that in the naked singularity case, the singularity $r=0$ is surprisingly given by the hyperbola
$- \tilde{t}^2 + \tilde{r}^2 = 1$ something which does not happen in the Schwarzschild case. 
In the latter case the singularity $r=0$ is on a very different zone known as zone $II$. Moreover, the
event horizon $r=2GM$ is given by the asymptotes of the fixed radial observes. 

In the present case of the naked singularity the asymptotes are at infinite past and infinity future
but do not represent any allowed value for coordinate $r$.

For instance $(\tilde{t}, \tilde{r})$ represents 2 dimensional Minkowski space-time. But in the naked singularity case 
instead of being divided in four zones as in the Schwarzschild
case, it is divided only in two parts, one for the original metric and one for the change of signature of
the metric. This is due to the fact that there is no event horizon in a naked singularity. 

The naked singularity we have presented here is an interesting toy model in order to 
see how naked singularities behave. And we have observed that the 2D naked singularity has its
own properties which are different to the four dimensional case.

\end{document}